\documentstyle[12pt]{article}
\newcommand{\pat}{\partial}
\newcommand{\be}{\begin{equation}}
\newcommand{\ee}{\end{equation}}
\newcommand{\bea}{\begin{eqnarray}}
\newcommand{\eea}{\end{eqnarray}}

\newcommand{\Mcal}{{\cal M}}

\newcommand{\Ocal}{{\cal O}}
\newcommand{\Rcal}{{\cal R}}

\newcommand{\half}{\frac{1}{2}}

\newcommand{\Lbar}{\bar{L}}
\newcommand{\wbar}{\bar{w}}
\newcommand{\lbar}{\bar{l}}

\newcommand{\bra}{\langle}
\newcommand{\ket}{\rangle}

\newcommand{\rhat}{\hat{r}}
\newcommand{\that}{\hat{t}}
\newcommand{\phihat}{\hat{\phi}}

\newcommand{\what}{\hat{w}}
\newcommand{\wbarhat}{\hat{\bar{w}}}
\newcommand{\newg}{\bar{g}_{n\omega}}

\newcommand{\ads}{SL$(2,\IR )_L \times$ SL$(2,\IR )_R$}

\def\IB{\relax\hbox{$\inbar\kern-.3em{\rm B}$}}
\def\IC{\relax\hbox{$\inbar\kern-.3em{\rm C}$}}
\def\ID{\relax\hbox{$\inbar\kern-.3em{\rm D}$}}
\def\IE{\relax\hbox{$\inbar\kern-.3em{\rm E}$}}
\def\IF{\relax\hbox{$\inbar\kern-.3em{\rm F}$}}
\def\IG{\relax\hbox{$\inbar\kern-.3em{\rm G}$}}
\def\IGa{\relax\hbox{${\rm I}\kern-.18em\Gamma$}}
\def\IH{\relax{\rm I\kern-.18em H}}
\def\IK{\relax{\rm I\kern-.18em K}}
\def\IL{\relax{\rm I\kern-.18em L}}
\def\IP{\relax{\rm I\kern-.18em P}}
\def\IR{\relax{\rm I\kern-.18em R}}
\def\IZ{\relax{\rm Z\kern-.5em Z}}
\begin{document}

\baselineskip 16pt

\begin{titlepage}
\begin{flushright}
CALT-68-2191  \\
hep-th/9808037 \\
August 1998
\end{flushright}

\vskip 1.2truecm

\begin{center}
{\Large {\bf Bulk and Boundary Dynamics  }} \\
{\Large {\bf in BTZ Black Holes}}
\end{center}

\vskip 0.8cm

\begin{center}
{\bf Esko Keski-Vakkuri}$^{1,2}$ 
\vskip 0.3cm
{\it California Institute of Technology \\
     Pasadena CA 91125, USA  \\
     email: esko@theory.caltech.edu}
\end{center}

\vskip 1.6cm

\begin{center}
{\small {\bf Abstract: }}
\end{center}
\noindent
{\small 
Recently, the AdS/CFT conjecture of Maldacena has been investigated
in Lorentzian signature by Balasubramanian {\em et. al.} We extend
this investigation to Lorentzian BTZ black hole spacetimes, and study
the bulk and boundary behaviour of massive scalar fields
both in the non-extremal and extremal case. Using the bulk-boundary
correspondence, we also evaluate the 
two-point correlator of operators coupling to the scalar field at 
the boundary of the spacetime, and find that it satisfies
thermal periodic boundary conditions relevant to the Hawking temperature
of the BTZ black hole. 
}
\rm
\vskip 2.2cm

\small
\begin{flushleft}
$^1$ Work supported in part by a DOE grant DE-FG03-92-ER40701.\\    
$^2$ Address after 1 September 1998: Department of Theoretical
Physics, Box 803, Uppsala University, S-75108 Uppsala, Sweden.
Email: esko@teorfys.uu.se 
\end{flushleft}
\normalsize 
\end{titlepage}

\newpage
\baselineskip 16pt

\section{Introduction}

There has been much  enthusiasm on the equivalence of string theory
on AdS$_{d+1} \times \Mcal $ spacetimes and conformal field theory
on the boundary of the anti-de Sitter space. The equivalence was conjectured
by Maldacena \cite{Mal}, and further 
developed for concrete computational tests 
by Gubser, Klebanov and Polyakov \cite{GKP} and by 
Witten \cite{Witten}. Two significant
aspects of the conjecture were quickly recognized and lot of activity has
happened around these themes. On the one hand, one can use the bulk theory 
(string theory or supergravity) to study various topics
in the boundary theory, such as correlation functions \cite{GKP,Witten,correl},
potentials between charged objects \cite{wilson} and so on. 
On the other hand, the
holographic aspect of the conjecture may realize the vision in the original
holography conjecture of t'Hooft and 
Susskind \cite{SHW,S}, \cite{Witten,SW}: that the holographic projection
to the boundary of the spacetime keeps track of all information which 
falls into
a black hole in the bulk and is thus never lost. 

The latter aspect has been studied by 
Maldacena and Strominger \cite{Str,Malstr}, who studied
BTZ black holes \cite{BTZ}, which have an asymptotic (local) 
anti-de Sitter 
geometry\footnote{There are also other black holes which have an asymptotic 
local anti-de Sitter geometry, {\em e.g.} the Schwarzschild black hole in
AdS space \cite{HP}
which is relevant for Yang-Mills theories at finite temperature
\cite{Witten,Witten2}.}.
The BTZ black hole is constructed from the AdS space by periodic 
identifications. It can be viewed as a Lorentzian orbifold, and as a simple
example of a more general class of Lorentzian orbifolds constructed from
AdS space, recently investigated in \cite{HorMar}. Strominger  
showed \cite{Str} how one can derive the Bekenstein entropy of the BTZ black
hole (see also the work of Birmingham {\em et. al.} \cite{BSS}) 
and, together with Maldacena \cite{Malstr},
also investigated other aspects relevant to the information problem.
In particular, they argued that the conformal field theory at the boundary
has a natural vacuum state analogous to the Minkowski vacuum in Rindler
space, which in the BTZ coordinates (analogous to the Rindler coordinates)
appears as a thermal multiparticle state. These kind of results are
necessary for understanding the black hole evaporation process purely
in terms of the boundary theory. We shall perform a related calculation in 
this paper: we evaluate the two-point function of boundary operators
in the BTZ coordinates, and show that it satisfies thermal periodicity
conditions relevant to the Hawking temperature of the BTZ black hole.

Despite this progress, a much more detailed understanding of the
conjecture and the equivalence between bulk states and boundary states is
needed before the information problem can fully be understood.
One step which would be helpful for the information problem would be a 
Hamiltonian version of Maldacena's conjecture, which would allow one to study 
dynamical aspects such as the time evolution of states. Originally, the
conjecture was formulated using a Euclidian signature, however a Lorentz 
signature version is needed for the Hamiltonian framework. Recently,
the conjecture was studied in Lorentz signature by Balasubramanian 
{\em et. al.} \cite{BKLT}, who laid out the general framework for the mapping
between the bulk field modes and boundary states. In Lorentz signature, the
former are divided in two classes of modes. The non-normalizable modes 
act as fixed backgrounds, and their 
boundary values act as sources for operators in the boundary theory. Turning
on a background non-normalizable mode thus changes the Hamiltonian of the
boundary theory. On the other hand, normalizable modes are quantized in the
bulk, and correspond to states in the Hilbert space of a given boundary 
Hamiltonian. 

In \cite{BKLT}, the bulk-boundary correspondence was studied in AdS using
two different coordinate systems. It would be
desirable to extend the work for black holes. In this paper, we  
extend the approach of \cite{BKLT} to 2+1 dimensional
BTZ black holes. In particular, we study
bulk scalar field modes in BTZ coordinates. The BTZ coordinates are analogous
to the Schwarzschild coordinates for ``ordinary'' black holes. They cover
the region outside the (outer) horizon. Schwarzschild modes play 
a central role in studies of dynamical black holes: the quantum state of
the black hole (the Kruskal vacuum) appears to be thermally excited in the
Schwarzschild coordinates, with temperature $T_H$. Thus it is an
important question to understand how these modes are 
understood from the boundary theory point of view. One immediate observation
is that BTZ black holes lose one powerful feature of AdS spaces.
The periodic identifications break the global \ads \ isometry group of
the AdS$_3$ space down to a $\IR \times$SO(2) subgroup. Thus, we lose the power
of the SL(2,$\IR$) representation theory in identifying bulk modes with
boundary states. The reduced global isometry group is something to keep in
mind when investigating BTZ black holes: although the spacetime is locally
AdS, some aspects of AdS spaces do not apply. 

The plan of the paper is as follows. In section 2 we quickly review the
key features and coordinate systems of 2+1 dimensional 
non-extremal and extremal BTZ black holes. 
In section 3 we compute the two-point function of boundary operators in 
BTZ coordinates acting on the Poincar\'{e} vacuum state. 
In section 4 we discuss their asymptotic conformal symmetries and global
isometries. In section 5 we solve the massive scalar field equations in
BTZ coordinates, first for non-extremal and then for extremal black holes,
and classify the mode solutions into non-normalizable and normalizable modes.
Section 6 is a brief summary.

We should note that solutions to the
massive scalar field equation in a non-extremal BTZ black hole spacetime 
have been studied before {\em e.g.} by Ichinose and Satoh \cite{modes} 
(see also \cite{SN})
and by Lee {\em et. al.} \cite{LKM,LMnew}. 
However, \cite{modes} contains 
only a partial list of
solutions, and all of the quoted references look at the solutions with
boundary conditions 
which are not directly applicable for the classification
of non-normalizable and normalizable modes within the
context of the Lorentzian BTZ/CFT conjecture.

\section{BTZ Coordinates}

AdS$_3$ is defined as the 
hyperboloid $-X_0^2-X_1^2+X_2^2+X_3^2 = -\Lambda^2$ embedded
in a space $\IR^{2,2}$ with 
metric $ds^2 = -dX_0^2-dX_1^2+dX_2^2+dX_3^2$\footnote{To make contact with
the notation in \cite{BKLT}, replace $X_0,X_1,X_2,X_3$ by 
$U,V,X,Y$.}. 
The induced
metric in the AdS$_3$ space can be written in global coordinates covering
the whole manifold, 
\be
 ds^2 = -(R^2 +\Lambda^2) dt^2 + \frac{\Lambda^2}{R^2 + \Lambda^2} dR^2
     + R^2 d\theta^2 \ ,
\label{global}
\ee
where the radial coordinate $R\ge 0$ and the angle coordinate is periodic
with
$0 \leq \theta \leq 2\pi$. The time coordinate
is also periodic with $0\leq t \leq 2\pi$, but one can go to
the covering space CAdS$_3$ with $t\in \IR$.
Another often used coordinate system is the Poincar\'{e} 
system, which covers only half of the AdS manifold. The metric is
\be
 ds^2 = \frac{\Lambda^2}{y^2} [dy^2 + dw_+ dw_-] 
\label{poinc}
\ee
with $y\ge 0,\ w_{\pm} = x_1 \pm x_0;\ x_1,x_0 \in \IR$.

The BTZ black hole is obtained from AdS$_3$ by periodic identifications. 
For this construction, it is useful to introduce a third coordinate system,
dividing the AdS space into three regions and defining coordinates
for each patch \cite{BTZ}. (For Penrose diagrams of AdS$_3$ in the three
different coordinate systems, see {\em e.g.} \cite{BKLT}.)
Since we are interested only in the region corresponding to the
exterior of the black hole, the relevant coordinate patch is given by 
\be
     \left. \begin{array}{ll} X_0 = \rhat \cosh \phihat  
      & \ \ X_1 = \sqrt{\rhat^2 - \Lambda^2} \sinh \that \\
      X_2 = \rhat \sinh \phihat & \ \ X_3 = \sqrt{\rhat^2 
                                            - \Lambda^2} \cosh \that 
      \end{array} \right.
\label{btzpatch}
\ee
with $\rhat \ge \Lambda,\ \phihat , \that \in \IR$.
The metric takes the form
\be
  ds^2 = -(\rhat^2 - \Lambda^2) d\that^2 + \frac{\Lambda^2 }
 {\rhat^2 - \Lambda^2} d\rhat^2 + \rhat^2 d\phihat^2  \ .
\label{btz}
\ee
This can be rewritten in a form where we can recognize a non-extremal
black hole. We make the coordinate transformation
\bea
      \rhat^2 &=& \Lambda^2 \left( \frac{r^2 - r^2_-}{r^2_+ - r^2_-} \right)
           \ \ (r^2 \ge r^2_+)  \\ 
 \left( \begin{array}{c} \that \\ \phihat \end{array} \right) 
 &=& \frac{1}{\Lambda} \left( \begin{array}{cc} r_+ & -r_- \\ -r_- & r_+ 
 \end{array} \right) \left( \begin{array}{c} t/\Lambda \\ \phi 
  \end{array} \right) 
\label{map}
\eea
and the metric becomes
\be
 ds^2 = -\frac{(r^2 - r^2_+)(r^2 - r^2_-)}{\Lambda^2 r^2} dt^2 
 + \frac{\Lambda^2 r^2 }{(r^2 - r^2_+)(r^2 - r^2_-)} dr^2
 + r^2 (d\phi - \frac{r_+r_-}{\Lambda r^2} dt)^2  \ .
\label{schw}
\ee
At this stage $\phi \in \IR$. The BTZ black hole is obtained 
through the periodic identification
\be
      \phi \sim \phi + 2\pi \ .
\label{period}
\ee
The parameters $r_{\pm}$ above then correspond to the radii of the outer
and inner horizons, and they are related to the mass $M$ and the angular
momentum $J$ of the black hole by 
\bea
       M\Lambda^2 = r^2_+ + r^2_- \\
       J\Lambda = 2 r_+r_- \ .
\eea 
Note that the periodic identification (\ref{period}) corresponds to 
the identification
\be
    (\that ,\phihat ) \sim (\that - 2\pi r_- / \Lambda ,
    \phihat + 2\pi r_+ / \Lambda )
\label{period2}
\ee
in the coordinates (\ref{btzpatch}). 
So in that coordinate system, for a generic rotating black hole,
the identification (\ref{period2}) acts on both coordinates
$\that$ and $\phihat$. If the angular momentum is
zero, $r_- =0$, and just the angular coordinate $\phihat$ is periodically
identified. The identification (\ref{period2}) corresponds to a Lorentz boost
in the $X_0-X_2$ and $X_1-X_3$ planes. 

We also note that the metric (\ref{btz}) can formally be related 
to the form (\ref{global}) in the
global coordinate system, by the analytic continuation
\be
    \that = it \ ; \ \phihat = i\theta \ ; \ \rhat = i R \  .
\label{analytic}
\ee
We can also relate it to the Poincar\'{e} metric 
with the coordinate transformation
\bea
   w_{\pm} &=& \frac{\sqrt{\rhat^2 - \Lambda^2}}{\rhat} e^{\phihat \pm 
    \that} \nonumber \\
  y &=& \frac{\Lambda}{\rhat} e^{\phihat} \ .
\eea
One can easily check that this transforms (\ref{poinc}) to the 
form (\ref{btz}).

The above construction works fine for the non-extremal case $r_+ > r_-$. In the
extremal case the coordinate transformation (\ref{map}) becomes 
singular, but the
metric (\ref{schw}) is still fine, reducing to the form 
\be
 ds^2 = -\frac{(r^2-r^2_0)^2}{\Lambda^2 r^2} dt^2 + \frac{\Lambda^2 r^2 }
    {(r^2 - r^2_0)^2} dr^2 + r^2 (d\phi - \frac{r^2_0}{\Lambda r^2} dt)^2 \ .
\ee
In this case we can relate the metric to the 
Poincar\'{e} form (\ref{poinc}) using the 
transformation
\bea
       w_- &=& \frac{\Lambda}{2r_0} e^{2r_0 (\phi -t)/\Lambda} \nonumber \\
       w_+ &=& \phi + t - \frac{\Lambda r_0}{r^2 - r^2_0} \nonumber \\
       y &=& \frac{\Lambda}{\sqrt{r^2 - r^2_0}} e^{r_0 (\phi -t)/\Lambda} \ .
\eea

\section{Two-point Functions}

There have been many studies of correlation functions in the boundary
theory using the bulk-boundary correspondence \cite{correl}. To our knowledge,
these investigations have mostly been performed in the Poincar\'{e} coordinate
system. For black hole applications, it would be of interest to examine
how the machinery works in the BTZ coordinates (\ref{schw}). As
an illustrative example we shall investigate the two-point
function of operators coupling to the boundary values of a massive
scalar field. Let us quickly
review some basic points. 
A massive scalar field $\Phi$ with a mass $\mu$
in an AdS$_{d+1}$ space with Euclidian signature has a unique solution
to its equation of motion. In Poincar\'{e} coordinates,
\be
 ds^2 = \frac{dy^2 + d\vec{x}_d^2}{ y^2} \ ,
\ee
the
solution behaves as $\Phi \rightarrow y^{2h_-} \Phi (\vec{x})$
near the boundary ($y\rightarrow 0$). The boundary field $\Phi_0(\vec{x})$
is of mass dimension $2h_-$ and acts as a source 
to an operator $\Ocal (\vec{x})$ of mass dimension $2h_+$ in the boundary
theory, with the parameters $h_{\pm}$ given by 
\be
   h_{\pm} = \frac{1}{4} (d \pm \sqrt{d^2 + 4\mu^2}) \ .
\label{hs}
\ee
On the other hand, the bulk field $\Phi (y,\vec{x})$ can be obtained from
the boundary field $\Phi_0(\vec{x})$ using the bulk-boundary Green's function.
We focus on $2+1$ dimensions and parametrize the boundary coordinates 
$\vec{x}$ by $w_{\pm}$, so the Poincar\'{e} metric takes 
the form (\ref{poinc}). Then the bulk field $\Phi$
is obtained from the boundary
values $\Phi_0$ by
\be
 \Phi (y,w_+,w_-) = \int dw'_+ dw'_-  K_P(y,w_+,w_-;w'_+,w'_-) 
 \Phi_0 (w'_+,w'_- ) 
\label{bulk}
\ee
where the bulk-boundary Green's function $K$ is
\be
 K_P(y,w_+,w_-;w'_+,w'_-) = c \left( \frac{y}{y^2 + \Delta w_+ \Delta w_- 
 } \right)^{2h_+} 
\label{green}
\ee
with $\Delta w_{\pm} = w_{\pm} - w'_{\pm}$. A scaling argument can be used
to see that 
\be
   \Phi \rightarrow y^{2h_-} \Phi_0 (w_+,w_-)
\label{scalew}
\ee
near the boundary. 

The above relations are given in the Euclidian signature. In the case
of a Lorentzian signature, the relation (\ref{bulk}) between the bulk
field and its boundary value becomes a bit more complicated. The bulk
field is no longer uniquely determined by its boundary value, but it is
determined up to the addition of a normalizable mode \cite{BKLT}.
The normalizable
modes have a subleading asymptotic behavior $y^{2h_+} \phi_0 (w_+,w_-)$
near the boundary, compared with the scaling (\ref{scalew}) of the 
non-normalizable modes. In our two-point function calculation, we will
be interested in the leading asymptotic behaviour near the boundary.
We will ignore the contribution from the non-normalizable
modes and work in the Lorentzian signature in the same manner as in 
the Euclidean signature.    

We would like to examine how these relations are modified
in the coordinate transformation to the BTZ coordinates (\ref{schw}),
\bea
 w_{\pm} &=& \sqrt{\frac{r^2 - r^2_+}{r^2-r^2_-}} \ e^{2\pi T_{\pm}u_{\pm}}
           \nonumber \\
 y &=& \sqrt{\frac{r^2_+-r^2_-}{r^2-r^2_-}} \ e^{\pi [T_+ u_+ +T_- u_-] }
\label{pointobtz}
\eea
where
\bea
 T_{\pm} &\equiv& \frac{r_+ \mp r_-}{2\pi \Lambda} \nonumber \\
 u_{\pm} &\equiv& \phi \pm t/\Lambda \ .
\eea 
On the other hand, for the two-point function calculation it is sufficient to 
simplify things and only examine the region
$r \gg r_{\pm}$ in the BTZ coordinates, where the coordinate 
transformation (\ref{pointobtz})
between the Poincar\'{e} and the BTZ coordinates reduces to the simplified form
\bea
     w_{\pm} &=& e^{2\pi T_{\pm} u_{\pm}} \nonumber \\
         y   &=& \sqrt{\frac{r^2_+ -r^2_-}{r^2}} e^{\pi [T_+ u_+ +T_- u_-]} \ .
\label{coordmap}
\eea
The Poincar\'{e} boundary coordinates $w_{\pm}$ are related to the 
BTZ boundary coordinates $u_{\pm}$ by an exponential transformation. The 
boundary coordinates $u_{\pm}$ are also involved in the transformation
between the radial coordinates $y$ and $r$. This fits nicely 
with what happens to the
scaling relation (\ref{scalew}). The boundary field $\Phi_0$ has a conformal
dimension $(h,\bar{h}) = (h_-,h_-)$, so it transforms as
\be
    \Phi_0 (w_+,w_-) = \left( \frac{dw_+}{du_+} \right)^{-h_-}
              \left( \frac{dw_-}{du_-} \right)^{-h_-} \Phi_0 (u_+,u_-)
\label{scalebdry}
\ee
under (\ref{coordmap}). Then, using (\ref{coordmap}) and (\ref{scalebdry}),
we see that the scaling relation (\ref{scalew}) becomes
\be
    \Phi \rightarrow r^{-2h_-} \Phi_0 (u_+,u_-) \ .
\label{scaleu}
\ee
Starting from (\ref{bulk}), and
using the coordinate transformation (\ref{coordmap}), the scaling
relation (\ref{scalebdry}), the scaling of $dw_{\pm}$, and $h_+ = 1-h_-$,
we find that the relation (\ref{bulk}) between the bulk field and the
boundary field becomes
\be
 \Phi (r,u_+,u_-) = \int du'_+du'_- \ K_{{\rm BTZ}}(r,u_+,u_-;u'_+,u'_-)
  \Phi_0 (u'_+,u'_-)
\label{bulk2}
\ee
where the bulk-boundary Green's function takes the form
\be 
 K_{{\rm BTZ}}(r,u_+,u_-;u'_+,u'_-) = c' \frac{\left( 
 \frac{r^2_+-r^2_-}{r^2} \right)^{h_+} e^{-2\pi h_+ 
  [T_+\Delta u_+ + T_- \Delta u_-]}}{\left\{ \frac{r^2_+ -r^2_-}{r^2}
 + \left( 1- e^{-2\pi T_+ \Delta u_+} \right)
  \left( 1- e^{-2\pi T_- \Delta u_-} \right) \right\}^{2h_+} }
\label{greenu1}
\ee
with $\Delta u_{\pm} = u_{\pm} -u'_{\pm}$. Note that this is manifestly
invariant under translations of boundary coordinates. To describe
a black hole, we also need to 
remember to take into account the periodic identification (\ref{period}). 
This is done using the method of images\footnote{See \cite{modes} for
a similar discussion for the bulk Green's function.}. We shift
the angle coordinate $\phi$ by integer multiples of $2\pi$ and add an
infinite sum in front of the Green's function. In terms of the null coordinates
$u_+,u_-$,  the final result becomes
\be 
 K_{{\rm BTZ}}(r,u_+,u_-;u'_+,u'_-) = c' \sum^{\infty}_{n=-\infty} 
 \frac{\left( 
 \frac{r^2_+-r^2_-}{r^2} \right)^{h_+} e^{-2\pi h_+ 
  [T_+\Delta u_+ + T_- \Delta u_+ +(T_+ + T_-) 2\pi n]}}
 {\left\{ \frac{r^2_+ -r^2_-}{r^2}
 + \left( 1- e^{-2\pi T_+ (\Delta u_+ + 2\pi n)} \right)
  \left( 1- e^{-2\pi T_- (\Delta u_- + 2\pi n) } \right) \right\}^{2h_+} } \ .
\label{greenu}
\ee

Next, we shall use these results to calculate (ignoring
overall coefficients) the two point correlator
$\bra \Ocal (u_+,u_-) \Ocal (u'_+,u'_-) \ket $.
We proceed as in \cite{Witten} and evaluate the surface integral
\be
 I(\Phi ) = \lim_{r_s \rightarrow \infty} \int_{T_s} du_+ du_- 
 \sqrt{h} \ \Phi (\hat{e}_r \cdot \nabla ) \Phi \ ,
\label{Ii}
\ee
where $T_s$ is the surface $r =r_s$, $h$ its induced metric (from the
BTZ metric (\ref{schw})) and $\hat{e}_r$ is the unit vector normal to the
surface. The induced area element and the radial projection
of the gradient are found to be
\be
  \sqrt{h} \hat{e}_r \cdot \nabla = \frac{r^2}{\Lambda} \frac{r}{\Lambda} 
 \pat_r \ .
\label{radder}
\ee
To evaluate the radial derivative of $\Phi$, we use the asymptotic form
of (\ref{bulk2}) as $r\rightarrow \infty$:
\be
 \Phi (r,u_+,u_-) \sim r^{-2h_+} \int du'_+ du'_- 
 \frac{e^{-2\pi h_+ [T_+\Delta u_+ + T_- \Delta u_-]}}
    {\left( 1-e^{-2\pi T_+\Delta u_+} \right)^{2h_+}
      \left( 1-e^{-2\pi T_-\Delta u_-} \right)^{2h_+} }
   \Phi_0 (u'_+,u'_-) \ .
\label{approxphi}
\ee
(For notational simplicity, from now on we will  
suppress the infinite sum over
the periodic images. It is included as in (\ref{greenu}).) 
The radial derivative becomes 
\be
 \sqrt{h} \hat{e}_r \cdot \nabla \Phi \sim (-2h_+) r^{2h_-} \int du'_+ 
 du'_- \ \cdots \ .
\label{help}
\ee
Finally, substituting (\ref{help}) into (\ref{Ii}) and using (\ref{scaleu}),
we end up with
\be
 I (\Phi) \sim \int du_+ du_- du'_+ du'_- \Phi_0 (u_+,u_-)
 \frac{e^{-2\pi h_+ [T_+ \Delta u_+ + T_- \Delta u_-]}}
{\left( 1-e^{-2\pi T_+\Delta u_+} \right)^{2h_+}
      \left( 1-e^{-2\pi T_-\Delta u_-} \right)^{2h_+} } 
 \Phi_0 (u'_+,u'_-) \ .
\ee
Thus, we can read off the two point correlator
\be 
 \bra \Ocal (u_+,u_-) \Ocal (u'_+,u'_-) \ket
  \sim
 \frac{e^{-2\pi h_+ [T_+ \Delta u_+ + T_- \Delta u_-]}}
{\left( 1-e^{-2\pi T_+\Delta u_+} \right)^{2h_+}
      \left( 1-e^{-2\pi T_-\Delta u_-} \right)^{2h_+} } \ .
\label{corrres}
\ee
This agrees with the result that would be obtained simply from starting
with the two-point correlator
\be
 \bra \Ocal (w_+,w_-) \Ocal (w'_+,w'_-) \ket \sim \frac{1}{(\Delta w_+
 \Delta w_-)^{2h_+}}
\label{vacuum}
\ee
and using the scaling relations for dimension $(h_+,h_+)$ operators
under the coordinate transformation $w_{\pm} \mapsto u_{\pm}$.
The physical interpretation of the result is in agreement with the proposal
of Maldacena and Strominger in \cite{Malstr}. The correlator (\ref{vacuum})
is the two-point function for dimension $(h_+,h_+)$ operators in a vacuum
state, the Poincar\'{e} vacuum\footnote{For a more detailed discussion
on relating the bulk states with boundary states in the
framework of quantized operators, see \cite{BKLT}.}. 
The coordinate transformation (\ref{coordmap})
induces a mapping of
the operators $\Ocal (w_+,w_-)$ to new operators $\Ocal (u_+,u_-)$ by
a Bogoliubov transformation. The new operators $\Ocal (u_+,u_-)$ see the
Poincar\'{e} vacuum state as an excited density matrix. We can see that
in fact they see the Poincar\'{e} vacuum state as a thermal bath of
excitations in the BTZ modes, since the two-point function (\ref{corrres})
can be seen to be periodic under the imaginary shifts \cite{HH}
\be
 (\Delta t ,\Delta \phi ) \mapsto
(\Delta t+i\beta_H , \Delta \phi + i\Omega_H \beta_H )
\ee
where $\beta_H$ is the inverse Hawking temperature $1/T_H$ and $\Omega_H$
is the angular velocity of the outer horizon,
\be
   T_H = \frac{r^2_+ -r^2_-}{2\pi \Lambda^2 r_+} \ ; 
 \ \Omega_H = \frac{r_-}{\Lambda r_+} \ .
\label{parameters}
\ee 
The angular velocity $\Omega_H$ can be interpreted as a chemical
potential. Similar 
periodicity conditions were also 
discussed in \cite{modes,LO} in
the context of Hartle-Hawking Green's functions in the bulk of BTZ spacetime. 
Incidentally, note that it is the Hawking temperature $T_H$ which appears
as the periodicity. There are two other temperature parameters as well,
the ``left'' and ``right'' temperatures $T_{\pm}$ which 
appear in the boundary coordinate transformation
(\ref{coordmap}). In \cite{Malstr}, the stress tensor components
$T_{\pm \pm}$ for the left movers
and the right movers were evaluated, and they turned out to depend on
the left and right temperatures. If one views the BTZ black hole as a part
of a D1-D5 brane system \cite{Mal, Malstr}, these are 
related to the temperatures of the
left and right moving excitations on the effective string. It is satisfying
that the Hawking temperature $T_H$ appears where it is expected, 
in the periodicity of the boundary Green's function. It still
remains to be better understood if the Poincar\'{e} vacuum really is the 
appropriate vacuum state for the system. For example, it would be
interesting to see if it was equal 
to the Hartle-Hawking vacuum.   

\section{Asymptotic Symmetries and Global Isometries}

In the remainder of the paper, we shall discuss the symmetries of the
BTZ spacetime and investigate the mode solutions to the massive scalar
field equation in BTZ backgrounds.

Brown and Henneaux discovered \cite{BH}
that AdS$_3$ has an asymptotic conformal symmetry
generated by two copies of the Virasoro algebra. In the global coordinates,
the asymptotic symmetry generators $l_n,\lbar_n$ 
take the form
\be
     l_n = i e^{inw} \ [ 2 \pat_w 
                - \frac{\Lambda^2 n^2}{R^2} \pat_{\bar{w}} -inR \pat_R ]
\label{vgen}
\ee
where $n \in \IZ$ and $w=t+\theta ,\wbar = t-\theta$. The generators $\lbar_n$
are obtained by exchanging $w$ with $\wbar$. The generators obey the 
Virasoro algebra
\be
 [l_m,l_n] = (m-n) l_{m+n} + \frac{c}{12} (m^3-m) \delta_{m+n,0} \ ,
\label{valg}
\ee
similarly for $\lbar_n$. Furthermore, $[l_m,\lbar_n]=0$. The central charge
$c$ was found to be $c=3\Lambda /2G$.

In addition to the infinite dimensional asymptotic symmetry group, the 
AdS$_3$ inherits the SO(2,2) $\cong$ SL$(2,\IR )_L \times$ SL$(2,\IR )_R$
global isometry group from the space $\IR^{2,2}$ through the embedding
$-X^2_0-X^2_1+X^2_2+X^2_3 = -\Lambda^2$. The generators of the group SO(2,2)
consist of the rotation generators
$$
    L_{ab} = X_a\pat_b - X_b\pat_a
$$
in the $ab=01,23$ planes, and the boost generators
$$
    J_{ab} = X_a\pat_b + X_b\pat_a
$$
in the $ab=02,03,12,13$ planes. The SL$(2,\IR)_L$ generators are the linear
combinations\footnote{Our convention is such that the generators will match
with those of \cite{BKLT} after the notational
replacement $X_0,X_1,X_2,X_3 \mapsto
U,V,X,Y$.}
\be
    L_1 = (-L_{01}+L_{23})/2 \ ; \ L_2 = (J_{12}-J_{03})/2
 \ ; \ L_3 = (J_{02}+J_{13})/2 
\label{combi1}
\ee
and the SL$(2,\IR)_R$ generators are 
\be
    \Lbar_1 = (-L_{01}-L_{23})/2 \ ; \ \Lbar_2 = (J_{12}+J_{03})/2
 \ ; \ \Lbar_3 = (J_{02}-J_{13})/2 \ .
\label{combi2}
\ee
The $L$ generators commute with the $\Lbar$ generators, and both sets
obey the commutation relations
\be
  [L_1,L_2] =-L_3 \ ; \ [L_1,L_3] = L_2 \ ; \ [L_2,L_3] = L_1 \ .
\ee
The global isometry algebra SL$(2,\IR)_L \times$SL$(2,\IR)_R$ is isomorphic
to the global conformal algebra generated by the Virasoro generators
$l_0,l_{\pm 1},\bar{l}_0,\bar{l}_{\pm 1}$. Hence one would like to construct
linear combinations $L_0,L_{\pm}$ out of $L_1,L_2,L_3$ which satisfy the
commutation relations
\be
   [L_0,L_{\pm}] = \mp L_{\pm} \ ; \ [L_+,L_-] = 2L_0 \ .
\label{algebra}
\ee

In the global coordinates (\ref{global}), it was found 
that the convenient linear combinations
are 
\be
   L_0 = iL_1 \ ; \ L_{\pm} = \pm L_2 +iL_3 \ \ ,
\label{globalcomb}
\ee
and the explicit representation of (\ref{algebra}) is
\bea
  L_0 &=& i\pat_w  \nonumber \\
  L_{\pm} &=&  i\frac{\sqrt{R^2 + \Lambda^2}}{2R}
   \ e^{\pm iw } \left\{ \frac{2R^2 + \Lambda^2}{R^2 + \Lambda^2}
       \pat_{w} - \frac{\Lambda^2}{R^2 + \Lambda^2} \pat_{\wbar}
    \mp iR \pat_{R} \right\} \ ,
\label{globalrep}
\eea
where $w=t+\theta , \wbar = t-\theta$. The $\Lbar$ generators are obtained
by exchanging $w \leftrightarrow \wbar$.

In Poincar\'{e} coordinates, the convenient linear combinations turned
out to be 
\be
   L_0 = -L_2 \ ; \ L_{\pm} = i(L_1 \pm L_3) 
\label{poinccomb}
\ee
with the explicit representations
\bea
 L_0 &=& -\half y\pat_y - w_+ \pat_+ \nonumber \\ 
 L_- &=& i\Lambda \pat_+ \nonumber \\
 L_+ &=& -\frac{i}{\Lambda} [w_+ y\pat_y + w^2_+ \pat_+ + y^2 \pat_-] 
\label{poincrep}
\eea
where $w_{\pm} = x_1 \pm x_0$. The $\Lbar$ generators are obtained
by exchanging $w_+ \leftrightarrow w_-$.

Note that although both cases represent the same algebra (\ref{algebra}), they
are different in one aspect. In the global coordinates, (\ref{globalrep})
gives a hermitian $L_0$, and $L_{\pm}$ are adjoint 
operators: $L^{\dagger}_+ = L_-$. (This corresponds to the convention in 
\cite{BF}.) Near the boundary ($R\rightarrow \infty)$, the 
generators $L_0,L_{\pm}$ reduce to the asymptotic symmetry generators 
$l_0,l_{\pm 1}$. Thus, $L_0,L_{\pm}$ represent the bulk extensions of the
generators of the global conformal transformations.
However, in Poincar\'{e} coordinates, $L_0$ is antihermitian, and
$L_{\pm}$ are not adjoint operators. The reason for the difference is
that in global coordinates $L_0 \sim L_1$, but in Poincar\'{e} coordinates
$L_0 \sim L_2$. One can see that only if $L_0 \sim L_1$, it is possible
to construct linear combinations of $L_{1,2,3}$ such that $L_0$ is hermitian
and $L_{\pm}$ are adjoint operators. The transformation from global
to Poincar\'{e} coordinates induces a linear transformation in 
the \ads \ algebra, the basis transforming according to (\ref{poinccomb})
and (\ref{globalcomb}). As a result, the hermiticity properties of the
generators are altered. The lesson is that one should be careful in 
discussing
the unitarity properties of the bulk and boundary states corresponding to 
the Poincar\'{e} modes. 

What is the representation of (\ref{algebra}) in the BTZ coordinates 
(\ref{btzpatch})? The easiest way to find the representation is to 
start from the global coordinate representation (\ref{globalrep}),
and use the analytic
continuation trick (\ref{analytic}). The result is
\bea
  L_0 &=& -\pat_{\what} \nonumber \\
  L_{\pm} &=& -\frac{\sqrt{\rhat^2 - \Lambda^2}}{2\rhat}
   \ e^{\pm \what} \left\{ \frac{2\rhat^2 - \Lambda^2}{\rhat^2 - \Lambda^2}
       \pat_{\what} + \frac{\Lambda^2}{\rhat^2 - \Lambda^2} \pat_{\wbarhat}
    \mp \rhat \pat_{\rhat} \right\} \ ,
\label{btzrep}
\eea
where $\what = \that + \phihat, \wbarhat =\that - \phihat$. Another way
to derive these is to use (\ref{btzpatch}) and 
to find suitable linear combinations of (\ref{combi1}). The
linear combinations are now  
\be
 L_0 = -L_3 \ ; \ L_{\pm} = L_1 \mp L_2 \ .
\label{btzcomb}
\ee
As in the Poincar\'{e} representation, $L_0$ is antihermitian, and $L_{\pm}$
are not adjoint operators. The linear transformation between the global
generators (\ref{globalrep}) and (\ref{btzrep}) can be found from 
(\ref{btzcomb}) and (\ref{globalcomb}). 

The asymptotic symmetry generators in the BTZ coordinates are equally
easy to find.
The BTZ metric (\ref{btz}) has the same asymptotic form as the AdS$_3$ metric
(\ref{global}), so the asymptotic generators $l_n,\lbar_n$ are 
obtained from (\ref{vgen}) by replacing $R,\theta$ with $\rhat ,\phihat$. 

The generators (\ref{btzrep}) correspond to the
global symmetries of the AdS$_3$
space. An important point however is that after the periodic identifications,
(\ref{period}),(\ref{period2}), the global 
isometry group is no longer \ads , but a subgroup.
In the BTZ spacetime the global Killing vectors 
are $\pat /\pat_t ,\ \pat /\pat_{\phi}$ 
or $\pat /\pat_{\that} , \pat /\pat_{\phihat}$. These
generate time translations and rotations. 
So the global isometry group is broken
to \IR$\times$SO(2). The corresponding generators above are $L_0$ and
$\Lbar_0$, while $L_{\pm},\Lbar_{\pm}$ no longer generate global
spacetime symmetry transformations. 
Thus, we can redefine the global isometry generators to be $-2iL_0$
and $-2i\Lbar_0$, these are hermitian and equal to the 
asymptotic Virasoro generators $l_0$
and $\bar{l}_0$. Note that the vacuum 
state $|0\ket$ of
the CFT has to respect the global symmetries of the bulk. In the AdS$_3$ case
this means the condition $l_n|0\ket =0$ for $n=0,\pm1$. 
However, after the periodic
identifications it is sufficient that $l_0|0\ket =0$ only. (Similarly for
$\lbar$'s.) 

\section{Mode Solutions}

In Lorentz signature, the field equations of massive scalar fields in AdS
backgrounds 
have both non-normalizable and normalizable solutions. Explicit forms of
normalizable solutions in AdS spacetimes have been worked out 
in \cite{AIS,BrFr,MT,BKLT}.
We will now study the mode solutions for massive scalar fields in non-extremal
and extremal BTZ black hole backgrounds. In the non-extremal case, the mode
solutions have been studied before in \cite{modes,LKM,LMnew} but 
these studies have
used different boundary conditions from the ones that we will adopt. 
Our goal is to classify the modes into non-normalizable and normalizable modes,
the former being fixed classical backgrounds that give source terms
in the boundary theory, while the latter are quantized and correspond to 
states in the boundary theory.

The scalar field equation is 
\be
   \Box \Phi = \mu^2 / \Lambda^2 \Phi \ ,
\label{feqn}
\ee
where $\Box$ is the d'Alembertian in the appropriate coordinate system,
and $\mu^2$ stands for the mass$^2$ and coupling to the Ricci scalar $\Rcal$,
\be
  \mu^2 \equiv m^2 + \lambda \Rcal = m^2 -\frac{6\lambda}{\Lambda^2} \ .
\ee
For example, $\lambda = 0 \ (1/8)$ for minimally (conformally) coupled scalars.
The mass term $m^2$ includes contributions from the Kaluza-Klein reduction,
if we have started from higher dimensions, {\em e.g.} BTZ$\times S^3 \times
\Mcal_4$ would be relevant for the D1+D5 system.

\subsection{Global Isometries}

In solving the equation (\ref{feqn}), we can make use of the global isometries
generated by $\pat /\pat_t ,\ \pat /\pat_{\phi}$ and make the separation
of variables
\be
    \Phi = e^{-i\omega t + in\phi} f_{n\omega } \ ,
\label{separate}
\ee
where $f_{n\omega}$ only depends on the radial variable. 
If we make the periodic identification (\ref{period}), $n$ is quantized
and takes integer values. 
We can now
immediately see that in the non-extremal case the solutions $\Phi_{n\omega}$ 
are eigenmodes of $L_0,\Lbar_0$.
Both eigenvalues are imaginary, which is expected since we obtained 
the generators by an euclidian continuation from the global generators 
(\ref{globalrep}) and they turned out to be antihermitian. 
Before the periodic identifications, the global
symmetry is \ads . Its representations in the context of 
tachyon solutions of
(\ref{feqn}) and states of SL(2,$\IR$) WZW models
have been discussed in \cite{SN}.
After the identifications, the global isometry group is
$\IR \times$SO(2) generated by $l_0 = 2i\pat_w$
and $\bar{l}_0 = -2i\pat_{\wbar}$. 
The modes belong to a representation labelled by
real $(l_0,\bar{l}_0)$ conformal weights $(h,\bar{h})$, with
\bea
      h &=& \omega -n/\Lambda 
    \nonumber \\
   \bar{h} &=& \omega +n/\Lambda 
    \ .
\eea

\subsection{The non-extremal BTZ black hole}

We first solve for mode solutions in the non-extremal black hole coordinate
system (\ref{schw}). In this case we can first follow the discussion
by Ichinose and Satoh \cite{modes}. The d'Alembertian is
\be
   \Box = -\frac{1}{r^2N^2} \left[ r^2 \pat^2_t 
 - \left( \frac{r^2}{\Lambda^2} - M\right) \pat^2_{\phi} + J\pat_t\pat_{\phi}
 \right] + \frac{1}{r} \pat_r (r N^2 \pat_r) 
\ee
where 
$$
      N^2 = \frac{(r^2-r^2_+)(r^2-r^2_-)}{\Lambda^2 r^2} \ .
$$
Next, we use the global isometries and separate the variables as 
in (\ref{separate}).
The function $f_{n\omega}$
depends on the radial coordinate, we make a coordinate transformation
and take it to be $v=r^2/\Lambda^2$. The field equation now reduces 
to the radial equation
\be
 f''_{n\omega} + \frac{\Delta'}{\Delta} f'_{n\omega} + \frac{1}{4\Delta^2} 
   \{ n(Mn-J\omega ) -\mu^2 \Delta -(n^2 - \omega^2 \Lambda^2) \} 
 f_{n\omega } = 0 \ ,
\label{radieq}
\ee
where 
\be
  \Delta \equiv (v- v_+)(v-v_-) = ( v-r^2_+/\Lambda^2)(v-r^2_- /\Lambda^2) \ .
\ee
We then substitute the ansatz
\be
   f_{n\omega}= (v-v_+)^{\alpha}(v-v_-)^{\beta} g_{n\omega} 
\ee
where the exponents $\alpha ,\beta$ are given by
\bea
  \alpha^2 &=& -\frac{1}{4(v_+-v_-)^2} \left( r_+ \omega - r_- n /\Lambda 
          \right)^2 \nonumber \\
 \beta^2 &=& -\frac{1}{4(v_+-v_-)^2} \left( r_- \omega - r_+ n /\Lambda 
          \right)^2  
\eea
and $g_{n\omega}$ is a function of a rescaled radial variable 
\be
  u = \frac{v-v_-}{v_+-v_-} \ .
\ee
Note that in general $\alpha ,\beta$ are imaginary. Using
the Hawking temperature $T_H$ and angular velocity of the outer
horizon $\Omega_H$, given by (\ref{parameters}),
$\alpha$ becomes
\be
   \alpha = \pm \frac{i}{4\pi T_H} \ (\omega - \Omega_H n) \ .
\ee
This maps the radial equation to the hypergeometric equation
\be
 u(1-u) g''_{n\omega} +\{ c - (a+b+1) u\} g'_{n\omega} -abg_{n\omega} =0 \ ,
\label{hypeq}
\ee
where the parameters are
\be
   \left. \begin{array}{l} a = \alpha + \beta + h_{\pm} \\
                           b = \alpha + \beta + h_{\mp} \\
                           c = 2\beta +1  \end{array} \right.
\label{abc}
\ee
with
\be
   h_{\pm} = \half (1\pm \nu) \equiv \half (1 \pm \sqrt{1+\mu^2}) \ .
\label{hpm}
\ee
Note that (\ref{hpm}) is the same parameter as (\ref{hs}) (with $d=2$).
Here the coordinate $u$ has the range $1\leq u \le \infty$, between
the outer horizon and the boundary at infinity.

The solutions of (\ref{hypeq}) depend on the parameter $\nu$, in particular
whether it is integer or not. Let us work out the solutions case by case.

\subsubsection{$\nu$ not integer}

To find which solutions could be normalizable and which ones cannot, let
us first investigate their behaviour near the boundary $u=\infty $.
In this region, the two linearly independent 
solutions for the radial modes are labelled by $h_{\pm}$, they are 
\bea
  f^{(+)}_{n\omega} &=& (u-1)^{\alpha} u^{-h_+-\alpha}
          F(\alpha+\beta +h_+, \alpha-\beta +h_+; 1+\nu; u^{-1}) 
  \nonumber \\
  f^{(-)}_{n\omega} &=& (u-1)^{\alpha} u^{-h_--\alpha}
          F(\alpha+\beta +h_-, \alpha-\beta +h_-; 1-\nu; u^{-1}) 
\eea
where we have used the radial variable $u$. $F$ is a hypergeometric function.
The asymptotic behaviour near the boundary is
\bea
  f^{(+)}_{n\omega } &\sim & u^{-h_+} \nonumber \\
  f^{(-)}_{n\omega } &\sim & u^{-h_-} 
\eea
Now recall $h_{\pm} = (1\pm \nu)/2$. Thus, 
for $\nu > 1$, $\Phi^{(+)}_{n\omega}$
is a candidate normalizable mode, and $\Phi^{(-)}_{n\omega}$ is the
non-normalizable mode coupling to operators of dimension $2h_+$ in 
the boundary theory. However, for $0< \nu < 1$, both modes are well
behaved at the boundary. The situation is therefore similar to the global
mode case in \cite{BKLT}.

Let us now examine the behaviour of the modes $\Phi^{(\pm)}$ close to the
outer horizon $u=1$.\footnote{ At this point a reader familiar with the global 
mode analysis of \cite{BKLT} might wonder why we are not analyzing the
two regions in the reverse order. This is because in \cite{BKLT} it was
more important that the modes were well behaved in the origin, whereas in
our case the behaviour at the boundary is more important. Note also the
difference with the absorption coefficient calculations. There the primary
boundary condition is that there are only infalling waves at the horizon;
the solution near the boundary will then be a superposition of a 
non-normalizable and a normalizable mode, see \cite{LMnew}. }
Using the linear transformation relations of the hypergeometric functions,
we find,
\bea
  f^{(\pm)}_{n\omega} &=&   A_{\pm} \ (u-1)^{\alpha} u^{\beta}
 F(\alpha+\beta+h_{\pm},\alpha+\beta+h_{\mp} ; 2\alpha+1;1-u) \nonumber \\
\mbox{} & & \ \ \  + B_{\pm} \ (u-1)^{-\alpha} u^{-\beta} 
 F(-\alpha-\beta+h_{\pm},-\alpha-\beta+h_{\mp} ; -2\alpha+1;1-u) \ ,
\label{sum}
\eea
where 
\bea
 A_{\pm} &=& \frac{\Gamma (1\pm \nu) \Gamma (-2\alpha )}
      {\Gamma (-\alpha -\beta +h_{\pm}) \Gamma (-\alpha +\beta +h_{\pm})}
 \nonumber \\
 B_{\pm} &=& \frac{\Gamma (1\pm \nu) \Gamma (2\alpha )}
      {\Gamma (\alpha +\beta +h_{\pm}) \Gamma (\alpha -\beta +h_{\pm})} \ .
\label{coef}
\eea
Note that unlike the global modes in \cite{BKLT}, in this case the
denominators of the two coefficients have no poles, since $\alpha ,\beta$
are imaginary. In fact, we can see that the coefficients, and hence
the two terms in (\ref{sum}) are complex conjugates. Hence, near the
horizon the modes have the behaviour
\be
  f^{(\pm)} \sim e^{i\theta_0} (u-1)^{\alpha} + e^{-i\theta} (u-1)^{-\alpha}
\ee
where the phase factor $\theta_0$ is given by the ratio of the two coefficients
(\ref{coef}).Introducing a ``tortoise coordinate'' 
$r_* = \frac{1}{4\pi T_H} \ln (u-1)$, we can write the 
normalizable modes $\Phi^{(+)}_{n\omega}$ in the form
\be
\Phi^{(+)}_{n\omega} \sim e^{-i\omega t +in\phi} \cos 
         [(\omega - \Omega_H n)r_* +\theta_0]
\label{planewaves}
\ee
where we recognize the solution as a superposition of infalling and
outgoing plane waves, with a relative phase shift $2\theta_0$.
It is then obvious that the radial component of the scalar current
vanishes near the horizon. This is the natural boundary condition to
satisfy. Ichinose and Satoh \cite{modes} 
were interested in the geometric entropy associated
with the scalar field and used a different boundary condition.
They introduced an ultraviolet cutoff and imposed Dirichlet boundary conditions
at a stretched horizon, thereby obtaining the correct logarithmic divergence
for the associated geometric entropy. This boundary condition would 
imply quantized frequencies, but is not the correct one to impose 
in the present setup. In our case the frequencies (and thus the conformal
weights $h,\bar{h}$) will remain continuous. We also remark 
that the modes in (\ref{planewaves}) are 
related to the BTZ Kruskal modes in the same 
manner as for Schwarzschild black holes the asymptotic ``out'' modes
are related to the Kruskal modes. If the scalar field is in a vacuum state
with respect to the Kruskal modes, the Hartle-Hawking vacuum, 
the modes (\ref{planewaves}) are thermally excited at the
BTZ Hawking temperature $T_H$. Note however that in section 3 
the two-point point function (\ref{corrres}) was evaluated at the Poincar\'{e}
vacuum. Thus it would be interesting to see what is the relation of the
Hartle-Hawking and Poincar\'{e} vacuum.

\subsubsection{$\nu$ a nonnegative integer}

For completeness, we investigate also the case $\nu$ a nonnegative integer,
not studied in \cite{modes}. 
To investigate the behaviour at the boundary, it is convenient
to map the equation (\ref{hypeq}) into a new one involving a 
coordinate $w = 1/(1-u)$ and a rescaled mode function $\bar{g}_{n\omega}
= w^{-a} g_{n\omega}$ where $a$ is the parameter given by (\ref{abc}).
The resulting equation is
\be
  w(1-w) \newg'' + [C-(A+B+1)w] \newg' -AB \newg =0 
\label{neweq}
\ee
where\footnote{We have chosen $b\ge a$.} 
\be
 \left. \begin{array}{l} A = \alpha + \beta + h_- \\
          B = -\alpha +\beta + h_- \\
          C = 2h_- = 1-\nu \end{array} \right.
\ee
The boundary now corresponds to $w=0$. Consider first the
case $\nu =0$. In the boundary region we have the
two linearly independent solutions
\bea 
    \newg^{(1)} &=& F(A,B;1;w) \\
    \newg^{(2)} &=& \ln w \ F(A,B;1;w) + \sum^{\infty}_{l=1} k_l w^l \ ,
\eea
with the coefficients
$$
 k_l \equiv \frac{(A)_l (B)_l}{(l!)^2} [\psi (A+l) -\psi (A) 
         +\psi (B+l) -\psi (B) -2\psi (l+1) + 2\psi (l)]  \ .
$$
Therefore, the asymptotic behaviour of the corresponding radial modes
$f^{(1,2)}_{n\omega}$ is found to be
\bea
    f^{(1)}_{n\omega} &\sim & \frac{1}{\sqrt{u}} \rightarrow 0 
       \nonumber \\
  f^{(2)}_{n\omega} &\sim & \frac{\ln u}{\sqrt{u}} \rightarrow 0  \ .
\eea
Both modes vanish at the boundary, and correspond to normalizable
modes. The situation is then similar to the case $0<\nu <1$.
If $\nu =1,2,\ldots$, the two linearly independent solutions are
\bea
  \newg^{(1)} &=& w^{\nu} \ F(A+\nu,B+\nu ;1+\nu ;w) \nonumber \\
  \newg^{(2)} &=& \ln w \ \newg^{(1)} + \sum^{\infty}_{n=1} k_{n,\nu} w^n
       - \sum^{\nu}_{n=1} l_n w^{\nu -n} \ ,
\eea
with the coefficients
\bea
 k_{n,\nu} &\equiv & \frac{(A+\nu )_n (B+\nu)_n}{(1+\nu)_n \ n!}
    [\psi (A+\nu+n) -\psi (A+\nu) \nonumber \\
 \mbox{} & & \ \ \ \ \ \ \ \ \ \ \ \ \ \ \ \ \ 
     \ \psi (B+\nu+n) -\psi (B+\nu) -\psi (\nu+1+n) +\psi (\nu +1) 
          -\psi (1) ] \nonumber \\
 l_n &\equiv & \frac{(n-1)!(-\nu)_n}{(1-A-\nu)_n (1-B-\nu)_n} 
\eea
where $(-\nu)_n \equiv (-\nu )(\nu -1) \cdots (-\nu +n-1)$. In particular,
one can check that the coefficient $l_n$ multiplying a term $w^0$ is 
nonvanishing. The asymptotic behaviour near the boundary can then be
found to satisfy
\bea
  f^{(1)}_{n\omega} &\sim & u^{-h_+} \\
  f^{(2)}_{n\omega} &\sim & u^{-h_-} \ .
\eea
Thus, $f^{(1)}$ is the candidate normalizable mode, and $f^{(2)}$ is the
candidate non-normalizable mode coupling to a boundary operator of
dimension $2h_+$.

Using the linear transformation relations, one can find the behaviour
of the mode $f^{(1)}$ near the horizon, $w \rightarrow \infty$.
After some algebra, one can see that it will again be a sum of two
complex conjugate terms, with the familiar form
\be
 f^{(1)}_{n\omega} \sim e^{i\theta_0} (u-1)^{\alpha} + e^{-i\theta_0}
  (u-1)^{-\alpha} \ .
\ee
Again, it is a superposition of infalling and outgoing waves, and the
radial component of the flux vanishes.

\subsection{The extremal BTZ black hole}

We now find the mode solutions in the extremal case. In this case the
two regular singular points of the radial differential 
equation at the two horizons
will merge, and the equation can be mapped to a confluent hypergeometric
differential equation. Let us start again from the radial equation 
(\ref{radieq}). For the extremal black hole, $J=M\Lambda$, and
$$
 \Delta = (v-v_0)^2 = (\frac{r^2}{\Lambda^2} - M/2) \ .
$$
We first define a 
variable $x= 1/(v-v_0)$, for which the radial 
equation (\ref{radieq}) takes the form
\be
  f'' + \frac{1}{4} (A^2 - Bx^{-1} -\mu^2 x^{-2} ) f = 0 \ ,
\label{simp}
\ee
with the parameters 
\bea
   A^2 &\equiv &  r^2_0 (\omega - \Omega_H n)^2 
   \nonumber \\
   B &\equiv & \Lambda^2 (\omega^2 - \Omega^2_H n^2) \ .
\eea
where $\Omega_H = 1/\Lambda$ is the angular velocity  and
$r_0 = \sqrt{M\Lambda^2/2}$  is the radius of the horizon. 

Next, we rewrite the equation (\ref{simp}) as a Whittaker equation,
using a new variable $y =  iAx$:
\be
 f'' + \left( -\frac{1}{4} + \frac{\kappa}{y} + \frac{\frac{1}{4}-\lambda^2}
{y^2} \right) f = 0 
\label{Whitt}
\ee
where the parameters are
\bea
 \kappa &\equiv & iB/A = i \frac{\Lambda^2}{r^2_0}(\omega + \Omega_H n)
    \nonumber \\
 \frac{1}{4} - \lambda^2 &\equiv & -\frac{1}{4} \mu^2  \ . 
\eea
The last equation 
gives $\lambda =\lambda_{\pm} \equiv \pm \sqrt{1+\mu^2} /2$.
Note that we could have flipped
the sign in the definitions of $y$ and $\kappa$. Thus, a convenient 
choice for the 
two linearly independent solutions of (\ref{Whitt}) is
\bea
 f^{(+)} &=& M_{\kappa,\lambda_+} (y) + M_{-\kappa,\lambda_+} (-y) \nonumber \\
 f^{(-)} &=& M_{\kappa,\lambda_-} (y) + M_{-\kappa,\lambda_-} (-y) \ , 
\eea
where $M_{\kappa ,\lambda} (y)$ is a Whittaker function \cite{AbSt}.
We chose both of the solutions to be real valued.
The radial functions have the following
asymptotic behaviour at the boundary
\bea 
 f^{(+)}_{n\omega} &\sim & r^{-2h_+}\nonumber \\
 f^{(-)}_{n\omega} &\sim & r^{-2h_-}\ .
\eea
$f^{(-)}$ blows up at the boundary and is the candidate non-normalizable
mode, whereas $f^{(+)}$ decays and is the candidate 
normalizable mode. Near the horizon $f^{(+)}$ has
an oscillatory behavior characterized by
\be
   M_{\kappa ,\lambda} (y) \sim e^{y/2}\  y^{-\kappa} \ .
\ee
The radial current vanishes since $f^{(+)}$ was chosen to be real.

\section{Summary}

We studied various features of the bulk-boundary correspondence
in BTZ black hole spacetimes. We evaluated the bulk-boundary Green's
function for massive scalar fields
in the region far from the outer horizon, and used it to calculate
the Poincar\'{e} vacuum two-point function for operators 
in the boundary theory which couple
to the boundary values of the scalar field in BTZ coordinates. 
The two-point function
satisfies thermal boundary conditions, signaling that the Poincare 
vacuum looks like a thermal bath at the Hawking temperature $T_H$ of the black
hole. 

Then, we reviewed symmetry properties of BTZ black holes and examined
mode solutions to the massive field equations in non-extremal and extremal
backgrounds. The goal was to classify the solutions into non-normalizable
and normalizable modes according to the Lorentzian version \cite{BKLT}
of Maldacena's conjecture. This is a technical step, but we expect the
knowledge of the modes to be useful in further studies of BTZ black holes,
especially because the normalizable modes are thought to correspond to
states in the Hilbert space of the boundary theory. They are analogous
to the Schwarzschild modes for the familiar three dimensional black holes.
Such modes play a central role in studies of black hole evaporation, so
the hope is that understanding them as states in the boundary theory will
be helpful for the discussion of the evaporation process purely in the boundary
theory, according to the holographic principle.  

\bigskip

\bigskip

\begin{center}
{\large {\bf Acknowledgements}}
\end{center}

I would like to thank Per Kraus for several useful discussions, and
U. Danielsson, E. Gimon, M. Gremm, M. Kruczenski,
J. Louko, and J.-G. Zhou for helpful comments.


\bigskip

\bigskip

\begin{thebibliography}{1234567890}

\bibitem{Mal} J. Maldacena, ``The large N limit of 
superconformal field theories
and supergravity'', hep-th/9711200.
\bibitem{GKP} S.S. Gubser, I.R. Klebanov and A.M. Polyakov, ``Gauge theory
correlators from non-critical string theory'', hep-th/9802109.
\bibitem{Witten} E. Witten, ``Anti-de Sitter space and holography'', 
hep-th/9802150.
\bibitem{correl} M. Henningson and K. Sfetsos, hep-th/9803251; W. Muck
and K. Viswanathan, hep-th/9804035, hep-th/9805145; D.Z. Freedman, 
S.D. Mathur, A. Matusis, and L. Rastelli, hep-th/9804058; H. Liu and 
A. Tseytlin, hep-th/9804083; T. Banks and M. Green, hep-th/9804170; 
G. Chalmers, H. Nastase, K. Schalm, and R. Siebelink, hep-th/9805105;
A. Chezelbash, K. Kaviani, S. Parvizi, and A. Fatollahi, hep-th/9805162;
S.N. Solodukhin, hep-th/9806004; S. Lee, S. Minwalla, M. Rangamani, and
N. Seiberg, hep-th/9806074; G. Arutyunov and S. Frolov, hep-th/9806216;
H. Liu and A.A. Tseytlin, hep-th/9807097; E. D'Hoker, D.Z. Freedman,
and W. Skiba, hep-th/9807098; D.Z. Freedman, S.D. Mathur, A. Matusis,
and L. Rastelli, hep-th/9808006. 
\bibitem{wilson} J. Maldacena, {\sl Phys. Rev. Lett.} {\bf 80} (1998) 
4859 (hep-th/9803002);
S.-J. Rey and J. Yee, hep-th/9803001; J. Minahan, hep-th/9803111;
A. Brandhuber, N. Itzhaki, J. Sonnenschein, and 
S. Yankielowicz, hep-th/9803137, hep-th/9803263, hep-th/9806158; 
A. Volovich, hep-th/9803220; 
M. Li, hep-th/9803252, hep-th/9804175; U.H. Danielsson and A.P. Polychronakos,
hep-th/9804141; Y. Imamura, hep-th/9806162; 
J.A. Minahan and N. Warner, hep-th/9805104; E. Gimon and M. Gremm, to appear.
\bibitem{SHW} C.R. Stephens, G. `t Hooft, and B.F. Whiting, ``Black hole 
evaporation without information loss'', {\sl Class. Quant. Grav.}
{\bf 11} (1994) 621 (gr-qc/9310006).
\bibitem{S} L. Susskind, ``The world as a hologram'', {\sl J. Math. Phys} 
{\bf 36} (1995) 6377 (hep-th/9409089).
\bibitem{SW} L. Susskind and E. Witten, ``The holographic bound in
anti-de Sitter space'', hep-th/9805114.
\bibitem{HP} S.W. Hawking and D. Page, ``Thermodynamics of black holes
in anti-de Sitter space'', {\sl Commun. Math. Phys.} {\bf 87} (1983) 577.
\bibitem{Witten2} E. Witten, ``Anti-de Sitter space, thermal phase transition,
and confinement in gauge theories'', hep-th/9803131. 
\bibitem{Str} A. Strominger, ``Black hole entropy from near-horizon 
microstates'', {\sl J. High Energy Phys.} {\bf 02} (1998) 014 
(hep-th/9712251).
\bibitem{Malstr} J. Maldacena and A. Strominger, ``AdS$_3$ black holes
and a stringy exclusion principle'', hep-th/9804085.
\bibitem{BTZ} M. Ba\~{n}ados, C. Teitelboim and J. Zanelli, 
``The black hole in three-dimensional space-time'', 
{\sl Phys. Rev. Lett.} {\bf 69} (1992) 1849 (hep-th/9204099);
M. Ba\~{n}ados, M. Henneaux, C. Teitelboim, and J. Zanelli, ``Geometry
of the 2+1 black hole'', {\sl Phys. Rev.} {\bf D48} (1993) 1506 
(gr-qc/9302012).
For a review, see {\em e.g.}
S. Carlip, ``The (2+1)-dimensional black hole'', gr-qc/9506079.
For higher-dimensional BTZ black hole solutions,
see S. \AA minneborg, I. Bengtsson, S. Holst, and P. Peld\'{a}n, 
{\sl Class. Quant. Grav.} {\bf 13} (1996) 2707; 
M. Ba\~{n}ados, {\sl Phys. Rev.} {\bf D57} (1998) 1068;
M. Ba\~{n}ados, A. Gomberoff and C. Mart\'{\i}nez, ``Anti-de Sitter space
and black holes'', hep-th/9805087.
\bibitem{HorMar} G.T. Horowitz and D. Marolf, ``A new approach to string
cosmology'', hep-th/9805207.
\bibitem{BSS} D. Birmingham, I. Sachs, and S. Sen, ``Entropy of 
three-dimensional black holes in string theory'', {\sl Phys. Lett.} 
{\bf B424} (1998) 275 (hep-th/9801019).
\bibitem{BKLT} V. Balasubramanian, P. Kraus, and A. Lawrence, ``Bulk vs.
boundary dynamics in anti-de Sitter spacetime'', hep-th/9805171;
V. Balasubramanian, P. Kraus, A. Lawrence, and S.P. Trivedi, ``Holographic
probes of anti-de Sitter spacetimes'', hep-th/9808017.
\bibitem{modes} I. Ichinose and Y. Satoh, ``Entropies of scalar fields 
on three dimensional black holes'', {\sl Nucl. Phys.} {\bf B447}
(1995) 340 (hep-th/9412144).
\bibitem{SN} M. Natsuume and Y. Satoh, ``String theory on three dimensional
black holes'', {\sl Int. J. Mod. Phys.} {\bf A13} (1998) 1229 (hep-th/9611041);
Y. Satoh, ``Study of three dimensional quantum black holes'', Ph.D. Thesis
at University of Tokyo, hep-th/9705209.
\bibitem{LKM} H.W. Lee, N.J. Kim and Y.S. Myung,
``Probing the BTZ black hole with test fields'', hep-th/9803227;
H.W. Lee and Y.S. Myung, ``Greybody factor for the BTZ black hole and
a 5d black hole'', hep-th/9804095.
\bibitem{LMnew} H.W. Lee and Y.S. Myung, ``Greybody factors in the
AdS$_3$/CFT correspondence'', hep-th/9808002.
\bibitem{HH} J.B. Hartle and S.W. Hawking, ``Path-integral derivation
of black hole radiance'', {\sl Phys. Rev.} {\bf D13} (1976) 2188.
\bibitem{LO} G. Lifschytz and M. Ortiz, ``Scalar field quantization on 
the 2+1 dimensional black hole background'', {\sl Phys. Rev.} 
{\bf D49} (1994) 1929 (gr-qc/9310008).
\bibitem{BH} J.D. Brown and M. Henneaux, ``Central charges in the canonical
realization of asymptotic symmetries: and example from three-dimensional 
gravity'', {\sl Commun. Math. Phys.} {\bf 104} (1986) 207.
\bibitem{AIS} S.J. Avis, C.J. Isham, and D. Storey, ``Quantum field
theory in anti-de Sitter space-time'', {\sl Phys. Rev.} {\bf D18} 
(1978) 3565.
\bibitem{BrFr} P. Breitenlohner and D.Z. Freedman, ``Positive energy in 
anti-de Sitter backgrounds and 
gauged extended supergravity'', {\sl Phys. Lett.} {\bf B115} (1982) 197;
``Stability in gauged extended supergravity'' {\sl Ann. Phys., NY} 
{\bf 144} (1982) 249.
\bibitem{MT} L. Mezinescu and P.K. Townsend, ``Stability at a local maximum
in higher-dimensional anti-de Sitter space and applications to supergravity'',
{\sl Ann. Phys., NY} {\bf 160} (1985) 406. 
\bibitem{BF} A.O. Barut and C. Fronsdal, ``On non-compact groups: II. 
Representations of the 2+1 Lorentz group'', {\sl Proc. Roy. Soc.} (London)
{\bf A287} (1965) 532.
\bibitem{AbSt} M. Abramowitz and I.A. Stegun, ``Handbook of mathematical
functions'', Dover (NY) 1965.
\end{thebibliography}
\end{document}